\documentclass[aps,prd,preprint,tightenlines,groupedaddress,showpacs]{revtex4}
\usepackage{epsf,epsfig,graphics,graphicx}
\begin{document}

\title{Constraining Evolution of Quintessence\\ with CMB and SNIa Data}

\author{Wolung Lee}
\email{leewl@phys.sinica.edu.tw}
\author{Kin-Wang Ng}
\email{nkw@phys.sinica.edu.tw}
\affiliation{Institute of Physics, Academia Sinica, Nankang, Taipei, Taiwan 11529, R.O.C.}

\date{\today}

\begin{abstract}

The equation of state of the hypothetical dark energy component, which 
constitutes about two thirds of the critical density of the universe, 
may be very different from that of a cosmological constant. Employing a 
phenomenological model, we investigate semi-analytically the constraints 
imposed on the scalar quintessence by supernovae observations, and by the 
acoustic scale extracted from recent CMB data. 
We show that a universe with a quintessence-dominated phase in the dark 
age is consistent with the current observational constraints. 
\end{abstract}

\maketitle

\section{Introduction}
Recent astrophysical and cosmological observations such as dynamical mass, 
Type Ia supernovae (SNe), gravitational lensing, and cosmic microwave 
background (CMB) anisotropies, concordantly prevail a spatially flat universe 
containing a mixture of matter and a dominant smooth component, 
which provides a repulsive force to accelerate the cosmic 
expansion~\cite{ctrig}. The simplest candidate for this invisible 
component carrying a sufficiently large negative pressure is a true 
cosmological constant. The current data, however, are consistent with 
a somewhat broader diversity of such a repulsive ``dark energy'' 
as long as its equation of state approaches that of the cosmological 
constant at recent epoch. 
A dynamically evolving scalar field $\phi$ called ``quintessence'' (Q) is 
probably the most popular scenario so far to accommodate the dark energy 
component. It is very interesting and fundamentally important to distinguish 
the Q field from the true cosmological constant case.

Many efforts have been put forth to reconstruct the scalar potential $V(\phi)$
from observational data based on various reasonable physical motivations. 
They include pseudo Nambu-Goldstone boson, inverse power law, exponential, 
tracking characteristics, oscillating feature, and others~\cite{qmods}.
Several attempts have been made to test different 
Q-models~\cite{many}. Nevertheless, it proves to be premature at this stage 
to perform a meaningful data fitting to a particular quintessence model, or
to differentiate between the variations. Reconstruction of $V(\phi)$ 
would likely require next-generation observations.  

Since the scalar potential of the Q-field is scarcely known, 
it is convenient to discuss the evolution of $\phi$ through its equation of 
state, $p_\phi = w_\phi \rho_\phi$. Physically, $-1\le w_\phi\le 1$, 
where the former equality holds for a pure vacuum state. 
Lately some progress has been made in constraining the behavior of 
Q field from observational data. A combined large scale structure,
SNe, and CMB analysis has set an upper limit on Q models with a constant
$w_\phi < -0.7$~\cite{bond,bean}, and a more recent analysis of CMB observations 
gives $w_\phi=-0.82^{+0.14}_{-0.11}$~\cite{bac}. Furthermore, the SNe data 
and measurements of the position of the acoustic peaks in the CMB anisotropy 
spectrum have been used to put a constraint on the present 
$w_\phi^0 \le -0.96$~\cite{cope}. The apparent brightness of the farthest
SN observed to date, SN1997ff at redshift $z\sim 1.7$, is consistent with 
that expected in the decelerating phase of the flat $\Lambda$CDM model with 
$\Omega_\Lambda \sim 0.7$~\cite{riess}, inferring $w_\phi= -1$ for $z<1.7$. 

Given the above observational constraints, one would like to know the 
possible role played by the Q field in the early universe. If the
dark energy is a pure cosmological constant or it has a constant equation of 
state of $\lesssim -0.7$, it would be quickly dominated by the matter for 
$z>0.3$. The situation may be very different for quintessence with a 
time-dependent $w_\phi$.
Here we will adopt a model-independent approach in which a phenomenological 
form for the time-dependent $w_\phi$ is assumed and then used to unfold 
the dynamics of the Q field up to the epoch of the last scattering surface. 
In particular, we focus on the CMB constraints that apply to such a generic 
quintessence (GQ) scenario in the hope of distinguishing the Q field from the 
true cosmological constant and the constant equation of state.

\section{Quintessential Cosmology}
Consider a flat universe in which the total density parameter of the universe 
today is represented by $\Omega_0 = \Omega_m^0+\Omega_r^0+\Omega_\phi^0 = 1$ 
with a negligible $\Omega_r^0$ and $\Omega_m^0/\Omega_\phi^0\sim 1/2$. 
Assuming a spatially homogeneous $\phi$ field, the evolution of the cosmic 
background is governed by
\begin{eqnarray}
\frac{d\varrho_\phi}{d\eta}&=& -3a{\mathcal H} \left(1+w_\phi\right)
                               \varrho_\phi, 
\label{bkg1} \\
\frac{d{\mathcal H}}{d\eta}&=& -{3\over2}a{\mathcal H}^2- 
\frac{a}{2}(w_r\varrho_r+w_\phi\varrho_\phi), 
\label{bkg2}
\end{eqnarray}
where the conformal time $\eta=H_0\int dt a^{-1}(t)$ with the 
scale factor $a$ and the Hubble constant $H_0$, 
we have used $(d\phi/dt)^2=(1+w_\phi)\rho_\phi$ and 
$V(\phi)=(1-w_\phi)\rho_\phi/2$, and rescaled the energy density of the 
$i$-th component by the reduced Planck mass $M_p\equiv (8\pi G)^{-1/2}$ and 
$H_0$, i.e. $\varrho_i \equiv \rho_i/(M_pH_0)^2$. Accordingly, 
the dimensionless Hubble parameter is given by
\begin{equation}
    {\mathcal H}^2\equiv \left(\frac{H}{H_0}\right)^2 = 
    \Omega_m^0~a^{-3}+\Omega_r^0~a^{-4}+\Omega_\phi~{\mathcal H}^2.
\end{equation}
Therefore, given a prescribed function of time for the equation of
state, $w_\phi(\eta)$,
the problem is reduced to solving a set of first-order coupled ordinary 
differential equations (\ref{bkg1}) and (\ref{bkg2}) with the initial 
conditions set at the present time and the time variable running backwards.

It is well-known that $w_\phi\neq 0$ after last scattering would result in 
a time-varying Newtonian potential which produces anisotropies 
at large angular scales through the integrated Sachs-Wolfe effect. 
If we are to constrain dynamical Q models, it will be useful 
to define an $\Omega_\phi$-weighted average~\cite{huey99}
\begin{equation}
\langle w_\phi \rangle = \int_{\eta_{\rm dec}}^{\eta_0}
\Omega_\phi(\eta) w_\phi(\eta) d\eta
\times \left( \int_{\eta_{\rm dec}}^{\eta_0} 
\Omega_\phi(\eta) d\eta \right)^{-1},
\end{equation}
where $\eta_0$ and $\eta_{\rm dec}$ are respectively the conformal time today 
and at the last scattering. 
It was shown~\cite{huey99,bean} that as far as the CMB anisotropy power
spectrum is concerned, the time-averaged $\langle w_\phi \rangle$ 
is well approximated by an effective constant $w_\phi$ for models in 
which the Q component is negligible at the last scattering.

\section{The Generic Quintessence Scenario}
\begin{figure}
\leavevmode
\hbox{
\epsfxsize=6.0in
\epsffile{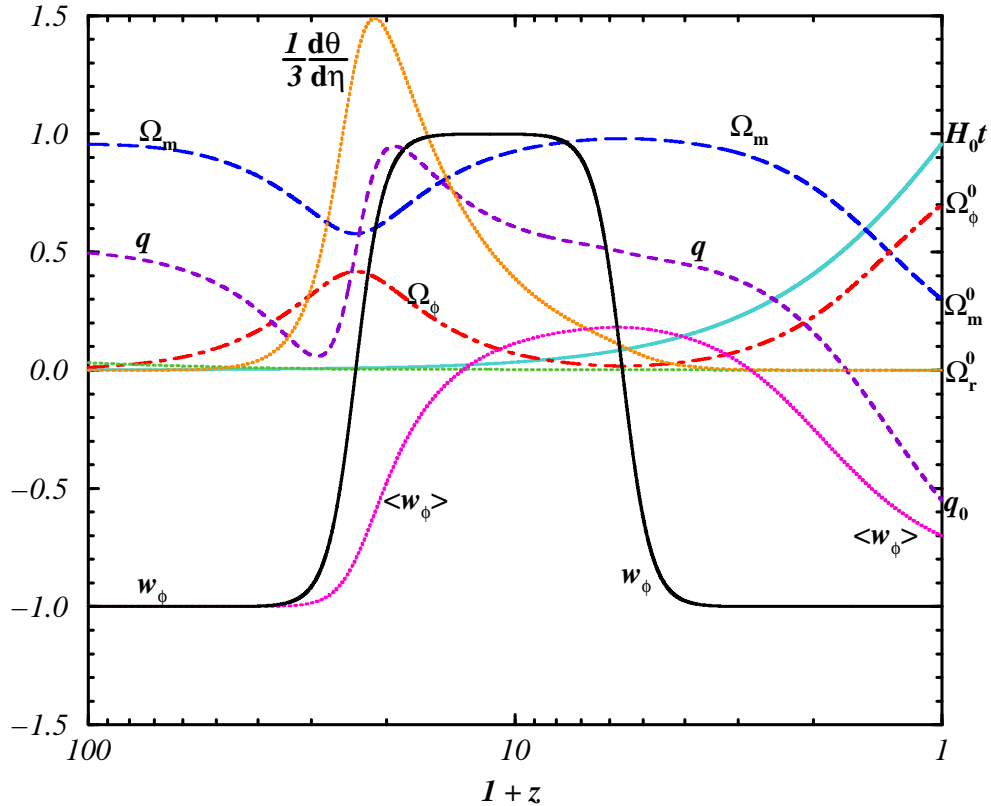}}
\caption{An example of GQ scenario}
\label{bg1}
\end{figure}

The generic quintessence scenario assumes a phenomenological form for the 
equation of state $w_\phi$ to accommodate as many observational outcomes 
as possible. Here we take $\Omega_m^0 = 0.3$ and 
$\Omega_r^0 = 9.89\times 10^{-5}$ with $w_r = 1/3$. 
Note that $\Omega_\phi^0 = 1-\Omega_m^0 - \Omega_r^0\simeq 0.7$.
Let us start with the form for $w_\phi$ at low redshifts.
In order to satisfy the above-mentioned SNe observational constraints
on $w_\phi$, we are bound to choose $w_\phi= -1$ for $z\lesssim 2$.
As such, the quintessence behaves just like a cosmological constant for
$0<z<2$, and it becomes dominated by the matter density at $z\sim 0.3$.
This can be seen in an example of the GQ in Fig.~\ref{bg1}. 
We have also plotted $d\theta/d\eta=a{\sqrt{(1+w_\phi){\varrho}_\phi}}$, 
where $\theta\equiv\phi/M_p$. For the Q field to obtain maximal dynamics
at low redshifts, $w_\phi$ has to increase abruptly to unity for $z>2$. 
However, if $w_\phi$ remains at unity for too long, it would induce an 
unacceptably large integrated Sachs-Wolfe effect on the CMB large-scale 
anisotropy. In Fig.~\ref{bg1}, we have chosen a width of the 
square-wave form of $w_\phi$ such that $\langle w_\phi \rangle\simeq -0.7$,
which saturates the above-mentioned upper limit from CMB data on a constant
$w_\phi$~\cite{bond}. 
We thus see that $\Omega_\phi$ can reach a value of $0.42$ at $z\sim 23$
at which $w_\phi$ is forced to decrease to a negative value.
On the other hand, a Q component with a constant $w_\phi= -0.7$ 
is dominated by matter for $z>0.5$.
Also note that at the present time $H_0 t=0.95$, and we have plotted the
history of the deacceleration parameter $q$.

The above GQ model exemplifies an extreme case on how the quintessence may be
different from a cosmological constant for $z>2$. If we shift the square-wave 
of $w_\phi$ to higher redshifts while keeping 
$\langle w_\phi \rangle\simeq -0.7$, 
the integrated Sachs-Wolfe effect on the CMB anisotropy power 
spectrum would be almost the same. Meanwhile,
the dynamics of the Q field will take place at higher redshifts accordingly.
This will be valid until the Q component becomes significant
at the last scattering.

\section {CMB Acoustic Peaks and Quintessence}
We now turn to study the effect of the presence of the Q component
at the time of last scattering on the CMB acoustic peaks.
The tightly coupled baryon-photon plasma experienced 
a serial acoustic oscillation just before the recombination epoch. 
The acoustic scale (the angular momentum scale of the acoustic oscillation) 
sets the locations of the peaks in the power spectrum of the CMB 
anisotropies~\cite{hu95}, and is characterized by
\begin{equation}
    l_A = \pi\frac{d_*}{h_s} = 
    \pi \frac{\eta_0-\eta_{\rm dec}}{\int_0^{\eta_{\rm dec}} c_s d\eta}~,
\label{la}
\end{equation}
where $d_*$ represents the comoving distance to the last scattering surface 
and $h_s$ denotes the sound horizon at the decoupling epoch with the sound 
speed $c_s$. Both $d_*$ and $h_s$ are affected in the presence of the Q 
component. Furthermore, the location of the $m$-th peak can be parametrized in 
practice by the empirical fitting formula, $l_m = l_A(m-\varphi_m)$, 
where the phase shift $\varphi_m$ caused by the plasma driving effect is 
solely determined by pre-recombination physics~\cite{hu01}. 
It was shown~\cite{doran2} that the shift of the third peak is relatively 
insensitive to cosmological parameters. Consequently, by assuming 
$\varphi_3 = 0.341$, the value of the acoustic scale derived from the 
analysis of BOOMERANG data~\cite{boom01} lies in the range~\cite{doran3}
\begin{equation}
    l_A = 316 \pm 8,
\label{bound}
\end{equation}
which is estimated to be within one percent error if the location $l_3$ 
is measured.

To impose the constraint from the peak separation of CMB, one needs to 
calculate the acoustic scale $l_A$ in Eq.~(\ref{la}) for different GQ models. 
Since the sound speed in the pre-recombination plasma 
is characterized by~\cite{hu95} 
\begin{eqnarray}
	c_s &=& \frac{1}{\sqrt{3(1+R)}}~~~{\rm with}~~\nonumber \\
	R&\equiv&\frac{3\rho_b}{4\rho_\gamma}\approx 30230\left(
	\frac{T_\gamma^0}{2.728K}\right)^{-4} \frac{\Omega_b^0 h^2}{1+z}~,
\end{eqnarray}
where $H_0=100 h {\rm km s^{-1} Mpc^{-1}}$ and 
the baryon-photon momentum density ratio $R$ sets the baryon loading 
to the acoustic oscillation of CMB, the sound horizon $h_s$ at the decoupling 
epoch can be determined by the differential equation
\begin{equation}
	\frac{dh_s}{d\eta}=c_s,
\end{equation}
coupled with the background evolution equations (\ref{bkg1}) and (\ref{bkg2}).
Using $\Omega_b^0 = 0.05$, $h = 0.65$, and 
fixing the last scattering surface at $z_{\rm dec} = 1100$, we have calculated 
the acoustic scales of different GQ models by shifting the square-wave 
of $w_\phi$ towards the last scattering surface while keeping 
$\left\langle w_\phi\right\rangle$ a constant.
\begin{figure}
\leavevmode
\hbox{
\epsfxsize=6.0in
\epsffile{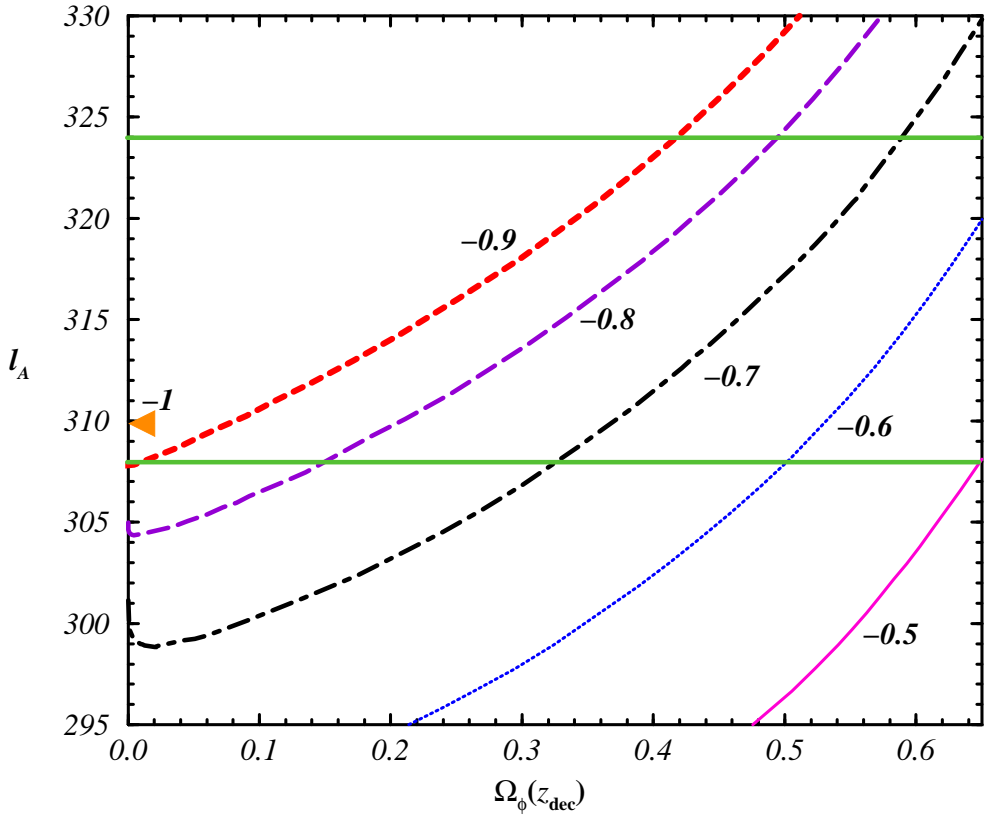}}
\caption{Acoustic scales of the GQ models are plotted as a function of the 
quintessence density at decoupling while keeping the $\Omega_\phi$-weighted 
averaged equation of state fixed. The model equivalent to the cosmological 
constant case has $l_A \simeq 310$ and is denoted by a solid triangle. 
The two horizontal lines signify the upper- and lower-bounds 
permitted by the BOOMERANG data.}
\label{qla}
\end{figure}
%
\begin{figure}
\leavevmode
\hbox{
\epsfxsize=6.0in
\epsffile{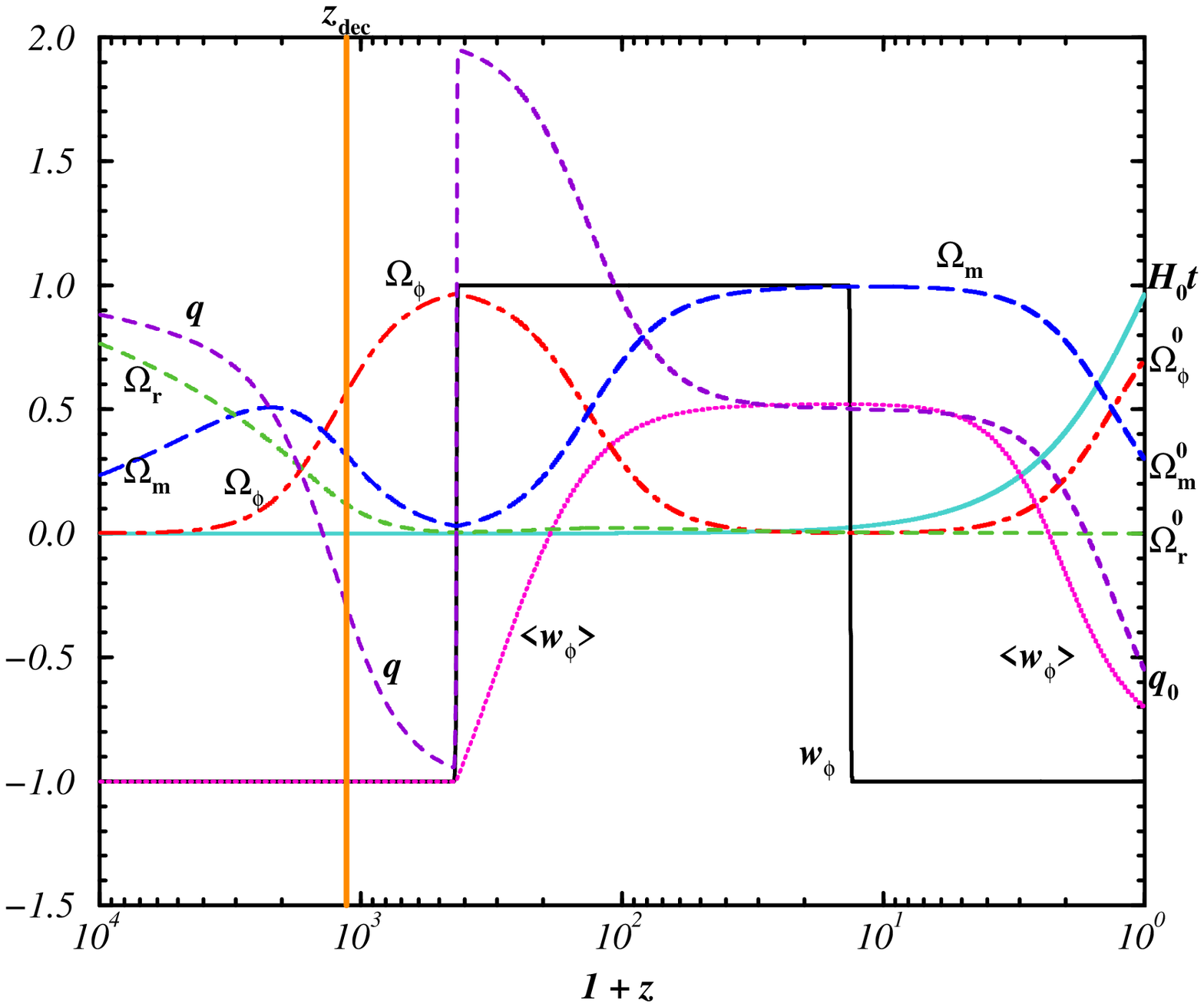}}
\caption{Background evolution of an extreme GQ model with 
($\left\langle w_\phi \right\rangle,~\Omega_\phi^{\rm dec}$) = ($-0.7,~0.59$). 
The last scattering surface is marked as the vertical 
line at $z_{\rm dec} = 1100$.}
\label{bg2}
\end{figure}
Figure~\ref{qla} plots the results against the Q energy density at 
decoupling $\Omega_\phi(z_{\rm dec})$ with contours
$\left\langle w_\phi\right\rangle = -0.5$, $-0.6$, $-0.7$, $-0.8$, $-0.9$, and 
$-1$. The last one is simply the $\Lambda$CDM model. In general, when 
$\left\langle w_\phi \right\rangle$ is fixed, the acoustic scale is 
proportional to the quintessence density at the last scattering surface.
On the other hand, for a fixed $\Omega_\phi(z_{\rm dec})$, 
the acoustic scale decreases as $\left\langle w_\phi \right\rangle$ increases. 
The two horizontal lines are drawn as the upper- and 
lower-bounds of the acoustic scale derived from the BOOMERANG 
data~(\ref{bound}). All GQ models lying in the region above the curve of 
$\left\langle w_\phi \right\rangle = -0.7$ and within the BOOMERANG bounds
are consistent with the current CMB data. Thus,
the model depicted in Fig.~\ref{bg1} is disfavored as far as the acoustic
peak location is concerned.
We have presented an extreme model located at the upper 
right corner of the allowed region in Fig.~\ref{bg2}, i.e.
($\left\langle w_\phi \right\rangle,~\Omega_\phi^{\rm dec}$) = ($-0.7,~0.59$). 
The figure shows the detailed background evolution of the model, in which 
the Q component dominates over the matter density for $1380>z>130$
and the deacceleration parameter $q$ is negative for $1400>z>430$.
\begin{figure}
\leavevmode
\hbox{
\epsfxsize=6.0in
\epsffile{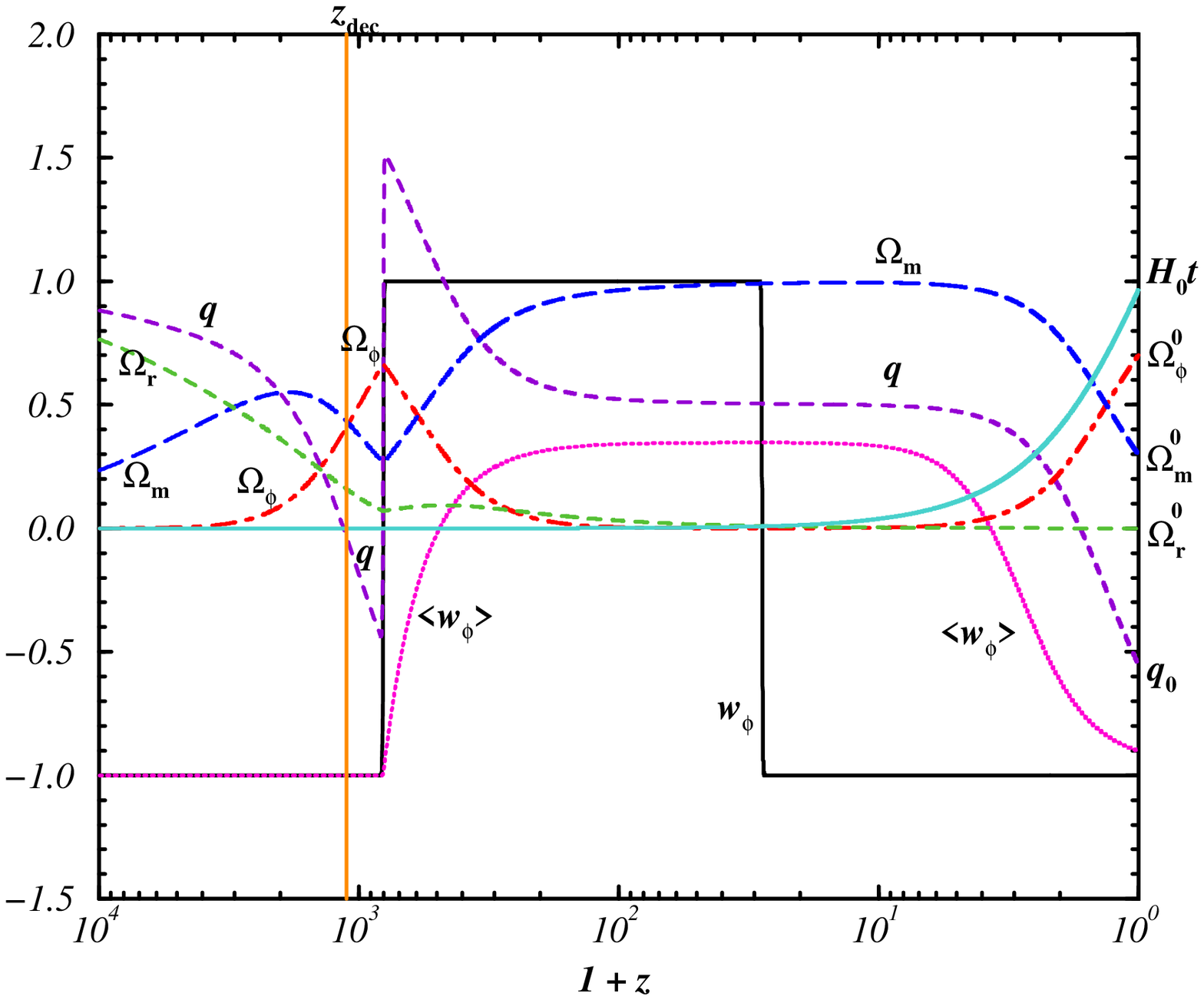}}
\caption{Background evolution of an extreme GQ model with 
($\left\langle w_\phi \right\rangle,~\Omega_\phi^{\rm dec}$) = ($-0.9,~0.42$).}
\label{bg3}
\end{figure}
%
\begin{figure}
\leavevmode
\hbox{
\epsfxsize=6.0in
\epsffile{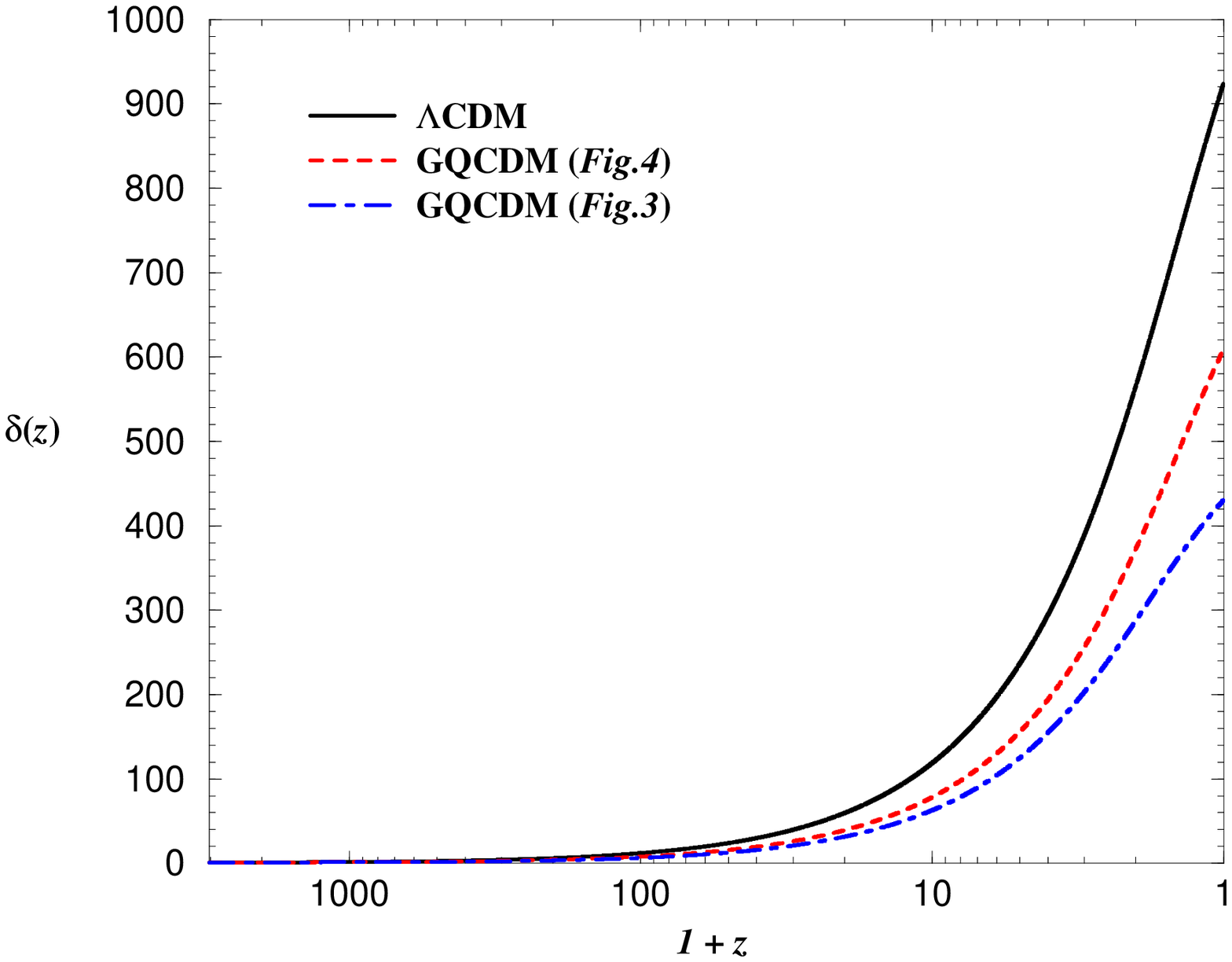}}
\caption{Growth functions of large-scale matter perturbations with 
unity normalization at radiation-matter equality time for the
$\Lambda$CDM model and the GQ models in Fig.~\ref{bg2} and Fig.~\ref{bg3}.}
\label{grow}
\end{figure}

\section{The Amplitude of Matter Power Spectrum}

It is well known that a non-zero $\Omega_\phi$ would lead to a suppression of
the growth of matter perturbations. It is therefore useful to study the 
constraint from the growth of structure on the GQ model. Let us consider 
the growth function for matter perturbations on scales smaller than the 
smoothing scale of the quintessence governed by the differential equation,
\begin{equation}
\frac{d^2\delta}{d\eta^2}+a{\mathcal H}\frac{d\delta}{d\eta}
-{3\over 2}a^2{\mathcal H}^2\Omega_m\delta=0.
\end{equation}
We have evaluated the growth function $\delta(\eta)$ for the extreme GQ model
in Fig.~\ref{bg2}. Compared to the $\Lambda$CDM model, the growth of 
$\delta$ from the radiation-matter equality to $\eta=\eta_0$ is suppressed by 
about a factor of 2 (see Fig.~\ref{grow}). When normalized to the COBE data, 
the resulting CMB anisotropy power spectrum has a significant excess of 
power at $l\sim 100$. 
In order to keep the suppression of the growth function stable
at about $30\%$ level, we have found that $\left\langle w_\phi\right\rangle$
has to be less than about $-0.9$ in Fig.~\ref{qla}. An accurate determination 
would require computing small-scale matter perturbations as well as 
performing a maximum-likelihood fitting to the existing CMB anisotropy data.
We have plotted in Fig.~\ref{bg3} an extreme model located at the upper end of 
the contour $\left\langle w_\phi\right\rangle= -0.9$ where 
$\Omega_\phi^{\rm dec}=0.42$, and the growth function of this model 
in Fig.~\ref{grow}. In this case, the Q component dominates over the matter 
density for $1100>z>590$ and the deacceleration parameter $q$ is negative for 
$1150>z>810$.

\section{Conclusion and Discussion}

We have investigated the evolution of the quintessence allowed by the
observational constraints from CMB and SNe, using a semi-analytic method with 
a simple square-wave function for the time-varying equation of state. 
Although the true equation of state, if there is any,
may be a complicated function of time, 
the square-wave should roughly capture the generic feature of the evolution 
of the quintessence. This generic quintessence model is sufficient for us
to confront the current observational data. Future high-precision data will 
tighten the constraints to this model and we may even need a more sophisticated 
model to parametrize the physics of the Q component.
Also, the present method gives more physical insights and 
is much simpler though less accurate than the numerically intensive maximum 
likelihood analysis of CMB data (see, e.g., Ref.~\cite{hansen}).

Three extreme GQ models have been presented. Figure~\ref{bg1}
shows the maximum dynamics that the Q field can attain at low redshifts for
$z>2$. The evolving Q field during the large-scale-struture formation
may have interesting cosmological implications. For instance, the authors in
Ref.~\cite{qpmf02} have attempted to generate primordial
magnetic fields from the dynamics of the Q field coupled to electromagnetism.
This electromagnetic Q field may also be responsible for the time-varying
fine structure constant ($\alpha$)~\cite{alpha} 
as it was recently claimed that the results of a
search for time variability of $\alpha$ using
absorption systems in the spectra of distant quasars
yield a smaller $\alpha$ in the past~\cite{webb}.
We have also studied the constraint from the growth of large-scale matter
perturbations on the GQ model. 
Figure~\ref{bg3} shows that the Q component can make up about
$40\%$ of the total energy density of the 
universe at last scattering. This result is consistent with the upper
bound $\Omega_\phi < 0.39$ during the radiation dominated epoch obtained by
performing a maximum likelihood analysis on the CMB data~\cite{hansen}.
In general, the GQ scenario bears a salient feature that the Q component 
overwhelms the matter during the dark age. It is worth studying in more details
about its influence on the evolution of matter perturbations and 
the subsequent structure formation.  
At last, we would like to point out that an acceleration of the universe in the
past is consistent with all observations. So the fact that the universe is 
accelerating today would not be quite unnatural.

\begin{acknowledgments}

This work was supported in part by the National Science Council, Taiwan, ROC
under the Grant NSC91-2112-M-001-026.

\end{acknowledgments}

\end{document}